\documentclass[final,5p,times,twocolumn]{elsarticle}

\usepackage{euscript,amssymb,amsmath}
\usepackage{amsfonts,latexsym}

\newcommand{\const}{\mbox{\rm const}\,}

\usepackage{graphicx}
\usepackage{epsfig}
\usepackage{color}

\newcommand{\be}[1]{\begin{equation}\label{#1}}
\newcommand{\ee}{\end{equation}}
\newcommand{\ba}[1]{\begin{eqnarray}\label{#1}}
\newcommand{\ea}{\end{eqnarray}}
\newcommand{\rf}[1]{(\ref{#1})}

\begin{document}

\begin{frontmatter}
		
\title{Effect of peculiar velocities on the gravitational potential in cosmological models with perfect fluids}
		
\author[a]{Alvina Burgazli}
\ead{aburgazli@gmail.com}
		
\author[b,c]{Valerii Shulga}
\ead{shulga@rian.kharkov.ua}
		
\author[d]{A.~Emrah Y\"{u}kselci}
\ead{yukselcia@itu.edu.tr}
		
\author[a]{Alexander Zhuk}
\ead{ai.zhuk2@gmail.com}

\address[a]{Astronomical Observatory, Odesa National University, Dvoryanskaya st. 2, Odesa 65082, Ukraine}

\address[b]{International Center of Future Science of the Jilin University, 2699 Qianjin St., 130012, Changchun City, China}

\address[c]{Institute of Radio Astronomy of National Academy of Sciences of Ukraine, 4 Mystetstv str., 61002 Kharkiv, Ukraine}
		
\address[d]{Department of Physics, Istanbul Technical University, 34469 Maslak, Istanbul, Turkey}

\begin{abstract} We consider a universe filled with perfect fluid with the constant equation of state parameter $\omega$. In the theory of scalar perturbations, we study the effect of peculiar velocities on the gravitational potential. For radiation with $\omega=1/3$, we obtain the expression for the gravitational potential in the integral form. Numerical calculation  clearly demonstrates  the modulation  of the gravitational potential by acoustic oscillations due to the presence of peculiar velocities. We also show that peculiar velocities affect the gravitational potential in the case of the frustrated network of cosmic strings with $\omega =-1/3$.	
\end{abstract}

\begin{keyword} cosmology \sep scalar perturbations \sep peculiar velocities \sep gravitational potential  \end{keyword}
		
\end{frontmatter}
	
\section{Introduction}
	
\setcounter{equation}{0}

Relativistic perturbation theory is a powerful tool for studying the formation of the large-scale structure of the Universe \cite{Bardeen,Peebles,Mukhanov,Mukhanov2,Durrer,Rubakov}. To model this process, usually, the matter is taken in the form of a perfect fluid, and its energy density fluctuations $\delta\varepsilon$ are the source of scalar perturbations (in particular, the source of the gravitational potential $\Phi$) in the perturbed Einstein equations \cite{Mukhanov2,Durrer,Rubakov}. 

However, the energy-momentum tensor may depend on metrics (see, e.g., formula (106.4) in the textbook \cite{Landau}). Therefore, the energy density fluctuation may explicitly depend on the gravitational potential. For example, in the case of a pressureless perfect fluid in the form of discrete point-like particles and its generalization to the continuous case we get $\delta\varepsilon=(\delta\rho +3\bar\rho\Phi)c^2$, where $\delta\rho$ and $\bar\rho$ are the fluctuation and averaged value of the mass density, respectively \cite{Mech1,Mech2}. Considering this fact, the cosmic screening approach was proposed in \cite{Ein1}, where the mass density fluctuations $\delta\rho$ are the source of scalar perturbations. Due to non-zero cosmological background $\bar\rho\neq 0$, the gravitational potential satisfies the Helmholtz-type equation and undergoes the exponential screening at a large distance from a massive source. An additional bonus of the cosmic screening approach is that, since the smallness of the contrast density is not assumed, it is valid both on the super-horizon and sub-horizon scales. It is worth noting that in \cite{Mukhanov2,Durrer,Rubakov}, the contrast of energy density of matter is considered in hydrodynamical approximation  $\delta\varepsilon/\varepsilon \ll 1$,  which means that such approach is valid at large cosmological scales.

Within the cosmic screening approach in \cite{Ein1,EinBril2,Duygu}, the pressureless dark and baryonic matter were considered in the form of point-like masses (e.g., galaxies and the group of galaxies). The exact expressions for the scalar and vector perturbations in the first order were found in 
\cite{Ein1}. They have a rather complicated form. In \cite{EinBril,EKZ1,EKZ2}, this model was generalized to the case of perfect fluids with linear and non-linear equations of state. It was found a system of equations for the scalar and vector perturbations. Unfortunately, the analytic expressions for them are absent, and equations should be solved numerically. 

Recently \cite{EE}, it was demonstrated that calculations could be considerably simplified if we combine both the cosmic screening and hydrodynamical approaches. That means, first, that we still consider the mass density fluctuations $\delta\rho$ as a source of the metric perturbations, and, second, we replace the peculiar velocities $v^i$ by the gradient of the peculiar velocity potentials $\partial_i v$:  $(\varepsilon+p)v^i\; \to \; (\bar\varepsilon+\bar p)\partial_iv$, where $p$ is the perfect fluid pressure. 
Such replacement leads to an insignificant loss of accuracy. First, the peculiar velocities do not play an essential role at small scales as the sources for the metric perturbations \cite{Ein1}, and, second, at large scales, the transitions to the hydrodynamical approximation is well-grounded. The prize for such a trick is the considerable simplification of the form of the gravitational potential \cite{EE}. It gives a possibility to see how peculiar velocities affect the screening length of the gravitational potential at large cosmological scales. That is an important point since the shape of the gravitational potential, in turn, affects the formation of structures in the Universe.
 
It is of interest to generalize such a combined approach for models where the matter source is different from dark matter. Because the Universe during its evolution passes through the radiation dominated stage with the equation of state parameter $\omega =1/3$, radiation is the most compelling case. The literature also addresses models with various equations of state. For this reason,  we start, for generality, with a model for an arbitrary $\omega =\mathtt{const} \neq -1/3$. Then we concentrate on radiation. In contrast to the pure cosmic screening approach considered in \cite{EinBril,EKZ1,EKZ2}, the combined approach makes it possible to obtain in the integral form expression for the gravitational potential produced by an individual inhomogeneity. In turn, this enables us to investigate how  acoustic oscillations modulate gravitational potential due to the presence of peculiar velocities.
We also discuss briefly the exceptional case $\omega =-1/3$ (e.g., the frustrated network of cosmic strings) and demonstrate here the effect of peculiar velocities on the gravitational potential.

\section{Setup of the model}

\setcounter{equation}{0}

We consider a universe filled with perfect fluid having the equation of state
\begin{equation}
\begin{aligned}
\label{2.1}
p=\omega \varepsilon\, ,\quad \omega =\const .
\end{aligned}
\end{equation}
For this model, the background Friedmann equation is
\begin{equation}
\label{2.2}
\frac{3\mathcal{H}^2}{a^2}=\frac{3 H^2}{c^2}=\kappa \bar \varepsilon\, ,
\end{equation}
where the Hubble parameter is $
H=(da/dt)/a=(c/a)(a'/a)\equiv (c/a)\mathcal{H}$, $c$ is the speed of light
and synchronous and conformal times are related as follows: 
$a d\eta=c dt$. Hereafter, the prime  denotes the derivative with respect to the conformal time $(\eta)$ and the overbar  indicates the background values. The constant $\kappa\equiv 8\pi G_N/c^4$ is introduced as well ($G_N$ is the Newtonian gravitational constant). Since the equation of state parameter $\omega$ is the constant, background equation of state is
\begin{equation}
\begin{aligned}
\label{2.3}
\bar p=\omega \bar\varepsilon\, .
\end{aligned}
\end{equation}
Therefore, from the energy conservation equation we get
\begin{equation}
\begin{aligned}
\label{2.4}
\bar\varepsilon = \frac{\bar A}{a^{3(1+\omega)}}\, ,
\end{aligned}
\end{equation}
where $\bar A$ is the constant of integration. Then, integration of the Friedman equation \rf{2.2} for $\omega\neq -1/3$ results in $a\sim \eta^{2/(1+3\omega)}\sim t^{2/[3(1+\omega)]}$ and $\eta \sim t^{(1+3\omega)/[3(1+\omega)]}$. For example, in the case of radiation $\omega=1/3$ where Eq. \rf{2.4} takes the form
\begin{equation}
\begin{aligned}
\label{2.5}
\bar\varepsilon = \frac{\bar A_{\mathtt{r}}}{a^{4}}\, ,
\end{aligned}
\end{equation}
we get  $a(\eta)=A_1 \eta\, ,\, (0 \leq \eta< +\infty),\, 
a(t)=\sqrt{2A_1 ct}$ and $ct = (1/2)A_1\eta^2$ where $A_1 \equiv \left(\kappa \bar A_{\mathtt{r}}/3\right)^{1/2}$. In the exceptional case $\omega = -1/3$, where Eq. \rf{2.4} is transformed into
\begin{equation}
\begin{aligned}
\label{2.6}
\bar\varepsilon = \frac{\bar A_{\mathtt{st}}}{a^{2}}\, ,
\end{aligned}
\end{equation}
integration of Eq. \rf{2.2} gives $a(\eta)=C e^{B\eta}\, , \, (-\infty <\eta< +\infty) \, ,$
where $C$ is the constant of integration, $a(t)=Bct$  and $ct=(C/B)e^{B\eta}$. Positive dimensionless constant $B \equiv \left(\kappa \bar A_{\mathtt{st}}/3\right)^{1/2} >0$.

Background matter is perturbed by inhomogeneities of perfect fluid. We consider only scalar perturbations. Then, in conformal Newtonian gauge perturbed metrics is \cite{Rubakov,Mukhanov}
\begin{equation}
\begin{aligned}
\label{2.7}
ds^2=a^2(\eta)[(1+2\Phi)d\eta^2-(1-2\Phi)d\textbf{r}^2]\, ,
\end{aligned}
\end{equation}
and the perturbed Einstein equations read
\begin{equation}
\begin{aligned}
\label{2.8}
\Delta\Phi-3\frac{a'}{a}\left(\Phi'+\frac{a'}{a}\Phi\right)=\frac12\kappa a^2\delta\varepsilon ,
\end{aligned}
\end{equation}
\begin{equation}
\begin{aligned}
\label{2.9}
\Phi'+\frac{a'}{a}\Phi=-\frac12 \kappa a^2(\bar\varepsilon+\bar p) v\, ,
\end{aligned}
\end{equation}
\begin{equation}
\begin{aligned}
\label{2.10}
\Phi''+3\frac{a'}{a}\Phi'+\left(2\frac{a''}{a}-\frac{a'^2}{a^2}\right)\Phi=\frac12 \kappa a^2\delta p\, ,
\end{aligned}
\end{equation}
where $v(\eta,\textbf{r})$ is the peculiar velocity potential. 
In the case of the constant parameter $\omega$, the perturbed energy density and pressure are related as
as follows:
\begin{equation}
\begin{aligned}
\label{2.11}
\delta p=\omega \delta\varepsilon\, .
\end{aligned}
\end{equation}
Therefore, the squared speed of sound is equal to the parameter of equation of state:
\begin{equation}
\begin{aligned}
\label{2.12}
u_{\mathtt{s}}^2=\frac{\delta p}{\delta \varepsilon} = \frac{\bar p}{\bar\varepsilon} =\omega\, .
\end{aligned}
\end{equation}
The energy density fluctuation can be expressed as follows \cite{EinBril,EKZ1,EKZ2}
\begin{equation}
\begin{aligned}
\label{2.13}
\delta\varepsilon=\frac{\delta A}{a^{3(1+\omega)}}+ 3(1+\omega)\, \bar{\varepsilon}\, \Phi\, ,
\end{aligned}
\end{equation}
where $\Phi$ is singled out.

We rewrite the perturbed Einstein equation in momentum space with the help of the Fourier transform: 
\begin{equation}
\begin{aligned}
\label{2.14}
F(\textbf{r})=(2\pi)^{-3/2}\int_{\mathbb{R}^3}d\textbf{k}e^{i\textbf{kr}}\Tilde{F}(\textbf{k}) \;.
\end{aligned}
\end{equation}
Hereafter the tilde denotes the Fourier transformed quantities. 
For example, Eq. \rf{2.8} reads:
\begin{equation}
\begin{aligned}
\label{2.15}
k^2\Tilde\Phi+3\frac{a'}{a}\Tilde{\Phi}'+3\frac{a'^2}{a}\Tilde\Phi=-\frac12 \kappa a^2\Tilde{\delta\varepsilon}\, .
\end{aligned}
\end{equation}
Then, the sum of Eq. \rf{2.15} times $\omega$ and transformed Eq.\ \rf{2.10} results in
\begin{equation}
\begin{aligned}
\label{2.16}
\Tilde{\Phi}''+3\frac{a'}{a}(1+u_{\mathtt{s}}^2)\Tilde{\Phi}'+k^2u_{\mathtt{s}}^2\Tilde\Phi=0\, ,
\end{aligned}
\end{equation}
where we took into account the relation
\begin{equation}
\begin{aligned}
\label{2.17}
2\frac{a''}{a}-\frac{a'^2}{a^2}(1-3\omega)=-\kappa a^2(\bar p-\omega\bar\varepsilon)=0\, .
\end{aligned}
\end{equation}
following from the Friedmann equations. The case of dust $u_{\mathtt{s}}=0$ was investigated in \cite{EE}. Therefore, we will not consider it in our paper. The physical wavelength corresponding to the momentum $k$ is $\lambda_k =a/k$ and the sound horizon is 
$\lambda_{\mathtt{s}}\equiv u_{\mathtt{s}}H^{-1}c$.

Since in the case $\omega\neq-1/3$ the scale factor behaves as $a(\eta) \sim \eta^{2/(1+3\omega)}$, then Eq. \rf{2.16} reads
\begin{equation}
\begin{aligned}
\label{2.18}
\Tilde{\Phi}'' +\frac{6(1+\omega)}{1+3\omega}\frac{1}{\eta}
\Tilde{\Phi}' +u_{\mathtt{s}}^2
k^2\Tilde\Phi=0
\end{aligned}
\end{equation}
with the general solution  (see, e.g., 53:3:7 in \cite{Atlas}) 
\begin{equation}
\begin{aligned}
\label{2.19}
\Tilde\Phi (\eta) = C_1\eta^{\nu}J_{-\nu}(u_{\mathtt{s}}k\eta) 
+ C_2\eta^{\nu}J_{\nu}(u_{\mathtt{s}}k\eta)\, ,\quad 
u_{\mathtt{s}}\neq 0\, ,
\end{aligned}
\end{equation}
where $J_{\nu}$ are Bessel functions and
\begin{equation}
\begin{aligned}
\label{2.20}
\nu = -\frac{5+3\omega}{2(1+3\omega)}\, .
\end{aligned}
\end{equation}

\

\section{Relativistic perfect fluid}

In this section, we consider the case of radiation in detail. Since for radiation $\omega=1/3$, the general solution \rf{2.19} reads
\begin{equation}
\begin{aligned}
\label{3.21}
\Tilde\Phi (\eta) = C_1\eta^{-3/2}J_{3/2}(u_{\mathtt{s}}k\eta) 
+ C_2\eta^{-3/2}J_{-3/2}(u_{\mathtt{s}}k\eta)\, .
\end{aligned}
\end{equation}
In the considered case, $\lambda_k=a/k = A_1 \eta/k$ and
$\lambda_{\mathtt{s}}= u_{\mathtt{s}}  A_1\eta^2$, and for the modes outside the horizon  $\lambda_k = A_1 \eta/k >>
\lambda_{\mathtt{s}}= u_{\mathtt{s}}  A_1\eta^2\; \Longrightarrow 
u_{\mathtt{s}}k\eta << 1$. Taking into account the properties of the Bessel functions \cite{Atlas,Handbook}
\begin{align}
	\sqrt{\frac{\pi}{2x}}\;J_{3/2}(x)&=j_1(x)=-\frac{\cos x}{x}+\frac{\sin x}{x^2}\, , \label{3.22}\\
	\sqrt{\frac{\pi}{2x}}\;J_{-3/2}(x)&=j_{-2}(x)=-\frac{\cos x}{x^2}-\frac{\sin x}{x}\, \label{3.23} ,
\end{align}
(where $j_{1,-2}$ are the spherical Bessel functions) and asymptotes for small $x$
\begin{align}
	J_{3/2}(x) &\to \frac{4}{3\sqrt{\pi}}\left(\frac{x}{2}\right)^{3/2} \to 0\,, \label{3.24} \\
	J_{-3/2}(x) &\to \frac{-1}{2\sqrt{\pi}}\left(\frac{x}{2}\right)^{-3/2} \to-\infty\, , \label{3.25}
\end{align}
we can easily see that the only non-falling mode is
\begin{equation}
\begin{aligned}
\label{3.26}
	\Tilde\Phi(\eta)&=\Phi_{(i)} 3\sqrt{\frac{\pi}{2}}\frac{1}{(u_{\mathtt{s}} k\eta)^{3/2}}J_{3/2}(u_{\mathtt{s}} k\eta) \\[2mm]
	&=-3\Phi_{(i)}\frac{1}{(u_{\mathtt{s}} k\eta)^2}\left[\cos(u_{\mathtt{s}} k\eta)-\frac{\sin(u_{\mathtt{s}} k\eta)}{u_{\mathtt{s}} k\eta}\right]
\end{aligned}
\end{equation}
in full agreement with \cite{Rubakov}.
This solution tends to the constant: $\Phi(\eta) \to \Phi_{(i)}$ outside the sound horizon $u_{\mathtt{s}}k\eta \to 0$. The r.h.s. of \rf{3.26} demonstrates that acoustic oscillations modulate the gravitational potential which we will show explicitly below.

Therefore, the Fourier transformed l.h.s. of Eq. \rf{2.9} is
\begin{equation}
\begin{aligned}
\label{3.27}
&\Tilde{\Phi}'+\frac{a'}{a}\tilde\Phi \\
&=
3\Phi_{(i)}\frac{1}{\eta}\left[
\frac{2}{(u_{\mathtt{s}}k\eta)^2}\left[\cos(u_{\mathtt{s}}k\eta)-\frac{\sin(u_{\mathtt{s}}k\eta)}{u_{\mathtt{s}}k\eta}\right]
+\frac{\sin(u_{\mathtt{s}}k\eta)}{u_{\mathtt{s}}k\eta}
\right] \\
&=
-\Tilde\Phi(\eta)\frac{1}{\eta}\left\{
2+\frac{(u_{\mathtt{s}}k\eta)^2\sin(u_{\mathtt{s}}k\eta)}{u_{\mathtt{s}}k\eta\, \cos(u_{\mathtt{s}}k\eta)-\sin(u_{\mathtt{s}}k\eta)}\right\}\, .
\end{aligned}
\end{equation}
We can use this formula to determine the peculiar velocity via Eq.\ \rf{2.9}. It shows that the peculiar velocity undergoes the acoustic oscillations.
On the other hand, substitution of this expression into Eq.\ \rf{2.15} eliminates the time derivative allowing us to define the Fourier transform of the gravitational potential
\begin{equation}
\begin{aligned}
\label{3.28}
	\Tilde{\Phi}=\dfrac{-\dfrac{\kappa}{2}\dfrac{\Tilde{\delta A_r}}{a^2}}{k^2
	-\dfrac{3}{\eta^2}
	\Bigg[2+\dfrac{(u_{\mathtt{s}}k\eta)^2\sin(u_{\mathtt{s}}k\eta)}{u_{\mathtt{s}}k\eta\, \cos(u_{\mathtt{s}}k\eta)-\sin(u_{\mathtt{s}}k\eta)}
	\Bigg]+\dfrac{a^2}{\lambda_{\mathtt{r}}^2}}
\end{aligned}
\end{equation}
where we used the Fourier transform of the energy density fluctuation \rf{2.13} for radiation
\begin{equation}
\begin{aligned}
\label{3.29}
\Tilde{\delta\varepsilon}= \frac{\Tilde{\delta A_{\mathtt{r}}}}{a^4}+4\frac{\bar A_{\mathtt{r}}}{a^4}\Tilde\Phi\,  .
\end{aligned}
\end{equation}
We also introduced the screening length
\begin{equation}
\begin{aligned}
\label{3.30}
\lambda_{\mathtt{r}}^2\equiv \left[\frac{2\kappa \bar A_{\mathtt{r}}}{a^4}\right]^{-1} \;.
\end{aligned}
\end{equation}
It can be easily seen that
\begin{equation}
\begin{aligned}
\label{3.31}
\frac{a^2}{\lambda_{\mathtt{r}}^2}=\frac{6}{\eta^2}\, 
\end{aligned}
\end{equation}
and $\lambda_{\mathtt{r}}=\lambda_{\mathtt{s}}/\sqrt{2}$. 

Since the combination $\Tilde{\Phi}'+(a'/a)\tilde\Phi$ is proportional to the peculiar velocity, it describes in \rf{2.15} the effect of the peculiar velocity on the gravitational potential.   If we neglect the contribution of peculiar velocity, then the matter density fluctuation is
\begin{equation}
\begin{aligned}
\label{3.32}
\frac{\Tilde{\delta A_{\mathtt{r}}}}{a^4}= -\frac{2}{\kappa a^2}\left(k^2+\frac{a^2}{\lambda_{\mathtt{r}}^2}\right)\Tilde \Phi\equiv -\frac{2}{\kappa a^2}f_1(k)\Tilde\Phi\, .
\end{aligned}
\end{equation}
On the other hand, the peculiar velocity contribution leads to the appearance of an additional $k$-dependent term: 
\begin{equation}
\begin{aligned}
\label{3.33}
\frac{\Tilde{\delta A_{\mathtt{r}}}}{a^4} &=
-\frac{2}{\kappa a^2}
\left[k^2
-\frac{3}{\eta^2}
\frac{(u_{\mathtt{s}}k\eta)^2\sin(u_{\mathtt{s}}k\eta)}{u_{\mathtt{s}}k\eta\, \cos(u_{\mathtt{s}}k\eta)-\sin(u_{\mathtt{s}}k\eta)}
\right]\Tilde\Phi \\
&\equiv-\frac{2}{\kappa a^2}f_2(k)\Tilde\Phi\, ,
\end{aligned}
\end{equation}
where we took into account the relation \rf{3.31}.

To obtain an explicit form of the gravitational potential, it is necessary to determine the expression for the matter density fluctuations $\Tilde{\delta A_{\mathtt{r}}}/a^4$. 
Now we consider the model where this fluctuation is a localized inhomogeneity in the form of the delta function
\begin{equation}
\begin{aligned}
\label{3.34}
\frac{\delta A_{\mathtt{r}}}{a^4} \quad \longrightarrow \quad \frac{Mc^2 \delta(\textbf{r})}{a^3}\, ,
\end{aligned}
\end{equation}
where $M$ is an effective mass of this inhomogeneity. With the help of the inverse Fourier transform 
we get
\be{3.35}
\frac{\Tilde{\delta A_{\mathtt{r}}}}{a^4} = \frac{Mc^2 (2\pi)^{-3/2}}{a^3}\, .
\ee

Now, we can describe preliminary some properties of the gravitational potentials in the presence or absence of the peculiar velocities. First, both \rf{3.32} and \rf{3.33} deep inside the horizon  $\lambda_k=a/k \ll \lambda_{\mathtt{r}}$ behaves as
\begin{equation}
\begin{aligned}
\label{3.36}
\frac{\Tilde{\delta A_r}}{a^4} \approx 
-\frac{2}{\kappa}\frac{k^2}{a^2}\Tilde\Phi\, .
\end{aligned}
\end{equation}
Such a large $k$ limit corresponds to the short distances from an inhomogeneity. Therefore, this formula demonstrates that at short distances the gravitational potential has the Newtonian behavior for both cases. 

Eq.\ \rf{3.32} shows that the gravitational potential satisfies the Helmholtz equation. Therefore, if we neglect the peculiar velocity, the gravitational potential created by individual inhomogeneity has the Yukawa
potential form  with characteristic length of interaction $\lambda_{\mathtt{r}}/a$. That is, at large distances (i.e., small $k$), for example, outside of the horizon $a/k \gg \lambda_{\mathtt{r}} \Leftrightarrow k\eta \ll 1$, the gravitational potential decreases exponentially with the exponent 
$a/\lambda_{\mathtt{r}}$. In the case of Eq.\ \rf{3.33} (i.e., in the presence of the peculiar velocities), the Helmholtz equation is corrupted because of the additional $k$-dependent term. Hence,  
the gravitational potential is not the Yukawa one. As we show below, the potential is approximately described by the Yukawa potential modulated by acoustic oscillations. Moreover, outside the horizon $a/k \gg \lambda_{\mathtt{r}} \Leftrightarrow k\eta \ll 1$ Eq.\ \rf{3.33} accepts the following form
\begin{equation}
\begin{aligned}
\label{3.37}
\frac{\Tilde{\delta A_{\mathtt{r}}}}{a^4}\approx -\frac{2}{\kappa a^2}\left(k^2+\frac32\frac{a^2}{\lambda_{\mathtt{r}}^2}\right)\Tilde \Phi\, .
\end{aligned}
\end{equation}
Therefore, the effect of the peculiar velocity results in the prefactor 3/2 in comparison with Eq.\ \rf{3.32}, and at large distances the gravitational potential decreases exponentially with the exponent 
$\sqrt{3/2}\, a/\lambda_{\mathtt{r}}$. 

Coming back to Eqs.\ \rf{3.32} and \rf{3.33}, we can express the gravitational potentials in the position space as
\begin{equation}
\begin{aligned}
\label{3.38}
	\Phi_i(\textbf{r}) &= -\frac{1}{(2\pi)^{3/2}}\frac{\kappa}{2a^2}\int_{\mathbb{R}^3}d\textbf{k}e^{i\textbf{kr}}\frac{\Tilde{\delta A_{\mathtt{r}}}}{f_i(k)}\\[2mm]
	&=-\frac{G_N}{c^2}\frac{M}{ar}\frac{2}{\pi} \int_0^{\infty}dk\; \frac{k\sin(kr)}{f_i(k)}
\, ,\quad i=1,2\, ,
\end{aligned}
\end{equation}
where we used the relation \rf{3.35}.
Obviously, in the case of the Helmholtz Eq.\ \rf{3.32}, the gravitational potential has the Yukawa form
\begin{equation}
\begin{aligned}
\label{3.39}
\Phi_1(r)= -\frac{G_N}{c^2}\frac{M}{ar}e^{-ar/\lambda_{\mathtt{r}}}\, .
\end{aligned}
\end{equation}
However, in the case of  Eq. \rf{3.33}, the integral \rf{3.38} can be calculated only numerically\footnote{We do the numerical calculations in Python by discretizing the integral in \rf{3.38}. Since we know the analytic formula for $\Phi_1$ \rf{3.39}, to determine the efficiency of the code we check the relative error $\epsilon \!=\! |x_a-x_n|/|x_a|$ where $x_n$ and $x_a$ are the numerical and the analytical solutions, respectively. We have found that the error is in the order of $10^{-3}$ at most in the steep region of the potential and it rapidly decreases in the flat region as expected, and takes the values in the order of $10^{-6}$. Regarding the pure numerical case $\Phi_2$, by using the same code, we only can compare two successive numerical solutions, say $x_n$ and $y_n$, as $\epsilon \!=\! |x_n - y_n|/|x_n|$ to estimate the relative error. In this case, the error is in the order of $10^{-8}$ at most.}. Since $a^2/\lambda_r^2=6/\eta^2$, the gravitational potential depends parametrically on the conformal time $\eta$.

\begin{figure}[!ht]
\centering

	\includegraphics[width=.9\linewidth]{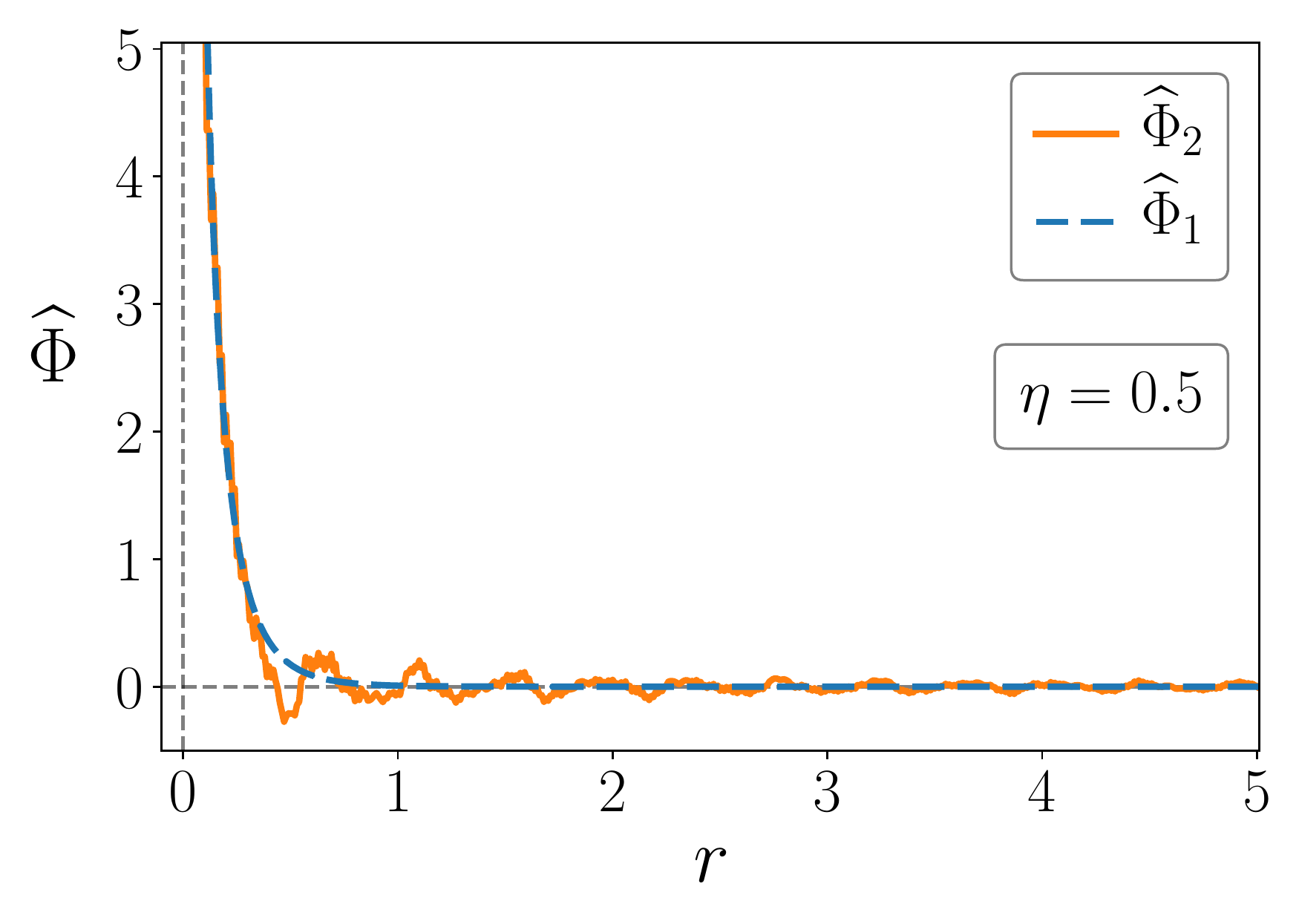}\\
	
	\includegraphics[width=.9\linewidth]{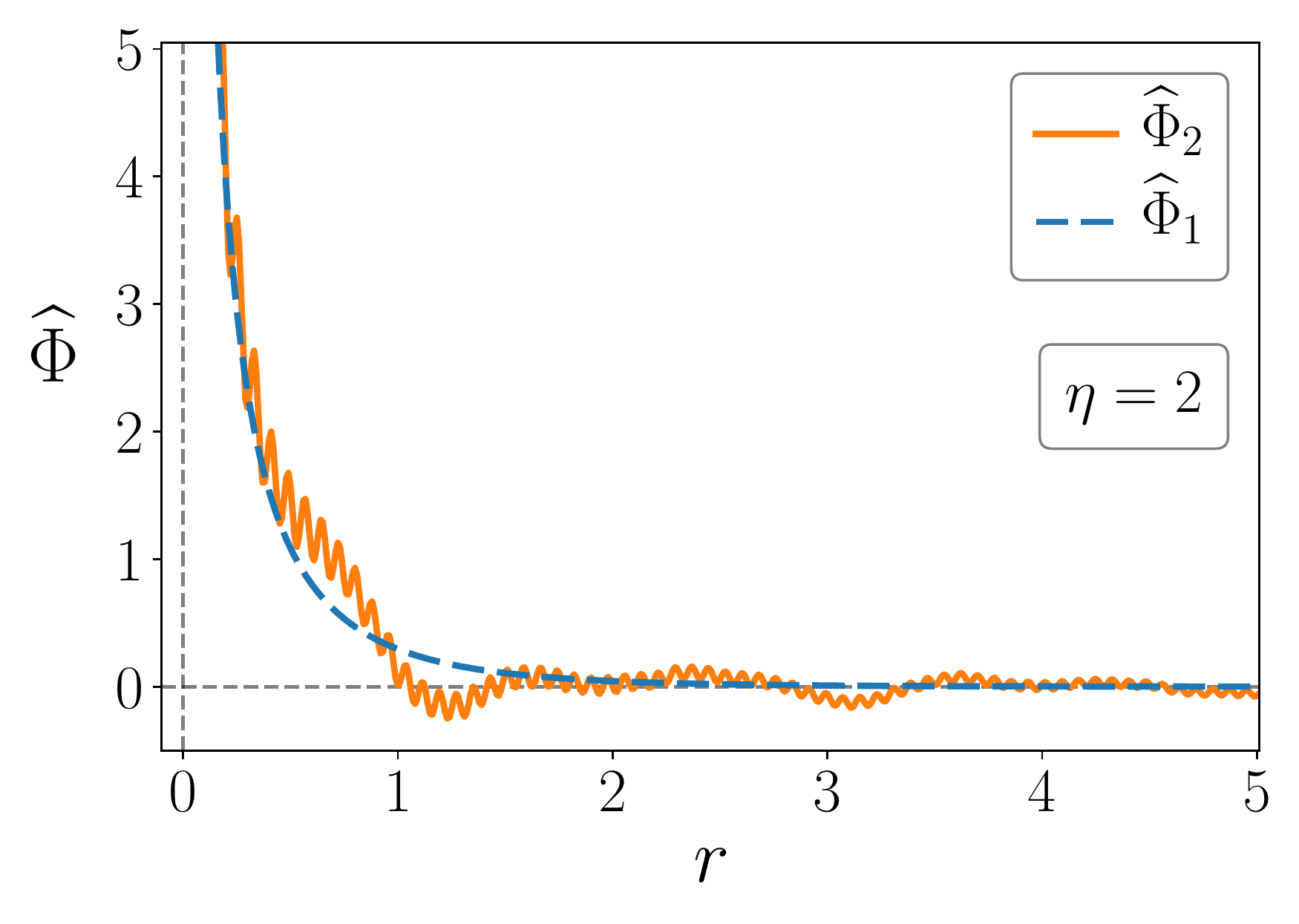}\\
	
	\includegraphics[width=.9\linewidth]{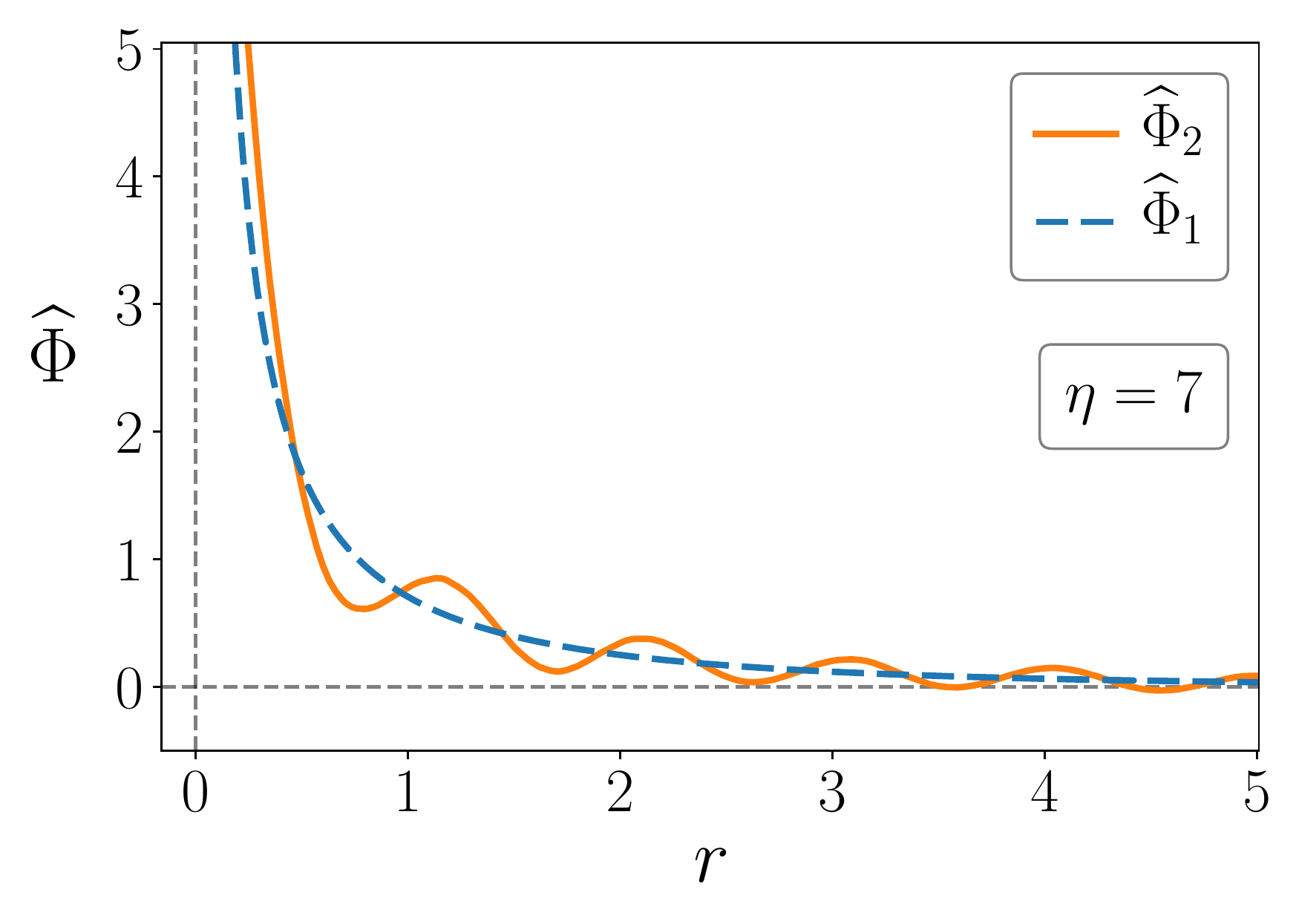}\\
	
	\vspace{-3mm}
	
\caption{Gravitational potentials in the case of radiation for different values of the parameter $\eta$. The dashed blue line corresponds to pure Yukawa potential and the solid orange line takes into account the effect of peculiar velocity.}
\label{fig.1}
\end{figure}

\newpage

In Figure \rf{fig.1}, we represent pictures for three different values of $\eta$ where the dimensionless potentials are defined as
\begin{equation}
\label{3.40}
\widehat{\Phi}_i(r)\equiv -\frac{c^2 a}{G_N M}\Phi_i\, , \quad i=1,2 \;.
\end{equation}

\begin{figure}[!ht]
 	\centering
	
	\includegraphics[width=.9\linewidth]{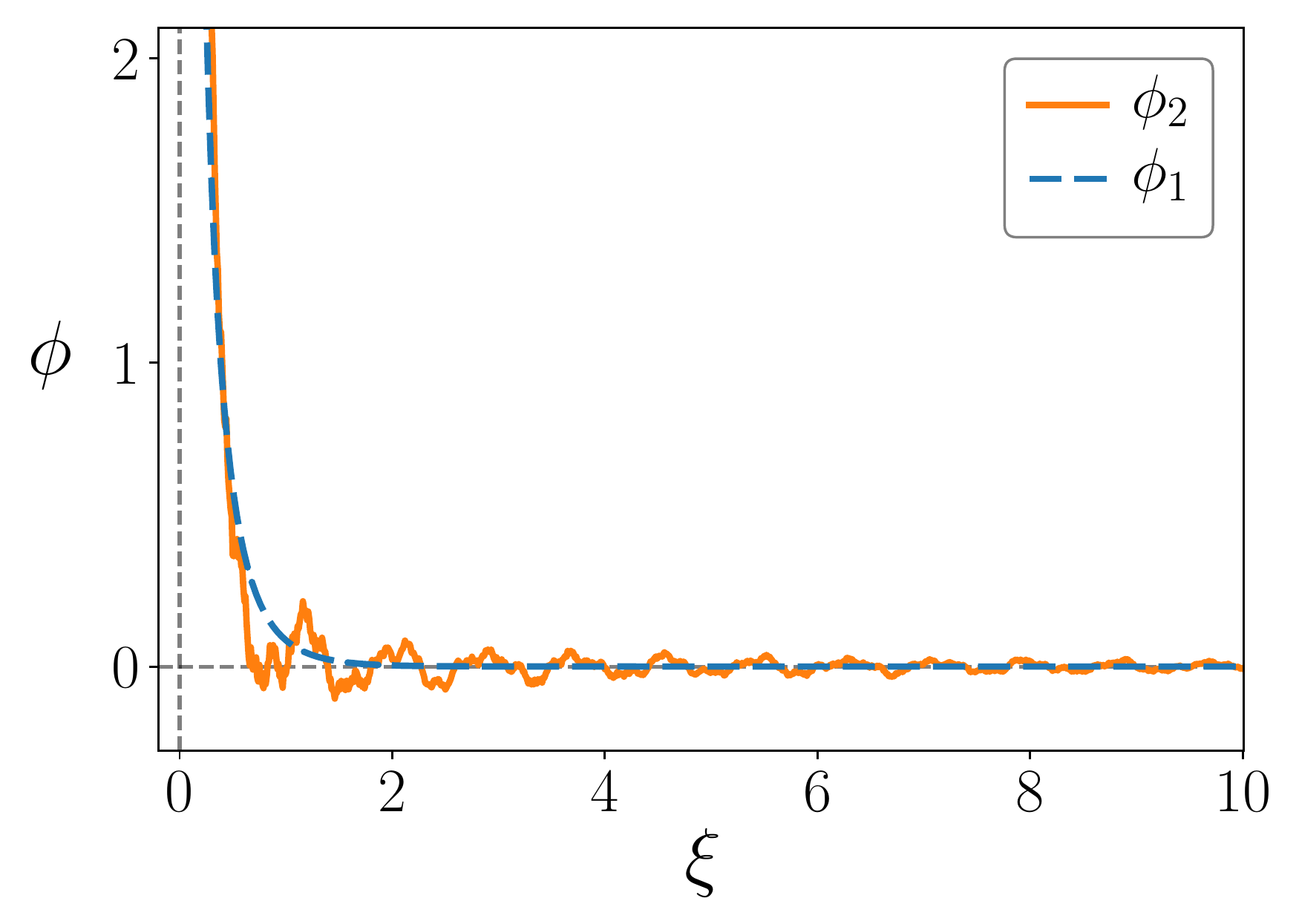} \\

 	\vspace{-3mm}

\caption{Gravitational potentials \rf{3.41} $\phi (\xi)$ where $\xi=r/\eta$. Dashed blue  and solid orange lines have the same meaning as in Figure \rf{fig.1}.}
\label{fig.2}
\end{figure} 

These pictures demonstrate the effect of peculiar velocities on the form of the gravitational potential (see the difference between $\widehat{\Phi}_1$ and $\widehat{\Phi}_2$), in particular, the modulation of the gravitational potential by acoustic oscillations in $\widehat{\Phi}_2$. 

It is also convenient to introduce new designations as $k\eta\equiv l$ and $r/\eta\equiv \xi$. Then, for dimensionless gravitational potentials we have
\begin{equation}
\begin{aligned}
\label{3.41}
\phi_i(\xi)\equiv -\frac{c^2 a\eta }{G_N M}\Phi_i=
\frac{1}{\xi}\frac{2}{\pi}
\int_0^{\infty}dl\; \frac{l\sin(l\xi)}{\widehat{f}_i(l)}
\, ,\quad i=1,2\, ,
\end{aligned}
\end{equation}
where
\begin{align}
\label{3.42n}
\widehat{f}_1(l) &\equiv l^2 +6\, ,\\
\label{3.}
\widehat{f}_2(l) &\equiv l^2 -
3\frac{(u_{\mathtt{s}}l)^2\sin(u_{\mathtt{s}}l)}{u_{\mathtt{s}}l\, \cos(u_{\mathtt{s}}l)-\sin(u_{\mathtt{s}}l)}\, .
\end{align}
Thus, for all values of $\eta$ we have only one figure depicted in Figure \rf{fig.2}.

\

\section{The frustrated network of cosmic strings: $\omega=-1/3$}

In this section, we briefly discuss the exceptional case $\omega=-1/3$. Such a model can describe the frustrated network of cosmic strings. Despite the negative sign of the speed of
sound squared, it was shown \cite{frust1,frust2} that
such a component could be stable if sufficiently rigid.

In the considered case, Eq. \rf{2.15} is
\begin{equation}
\begin{aligned}
\label{4.44}
\Tilde{\Phi}'' +2B\Tilde{\Phi}' -\frac{k^2}{3}\Tilde\Phi=0\, 
\end{aligned}
\end{equation}
with the general solution
\begin{equation}
\begin{aligned}
\label{4.45}
\Tilde\Phi = C_1 e^{-(B+\sqrt{B^2+k^2/3}\, )\eta} +C_2  e^{(-B+\sqrt{B^2+k^2/3}\, )\eta}\, .
\end{aligned}
\end{equation}
Obviously, the first solution is the decreasing one and we neglect it. 
Hence, 
\begin{equation}
\begin{aligned}
\label{4.46}
\Tilde\Phi = \Phi_i e^{(-B+\sqrt{B^2+k^2/3}\, )\eta}
=\Phi_i \left(\frac{B}{C}ct\right)^{-1 +\sqrt{1+k^2/(3B^2)}}\, ,
\end{aligned}
\end{equation}
where $\Phi_i$ is the value of $\Tilde\Phi$ at $\eta=0$.
Therefore, the l.h.s. of Eq.\ \rf{2.9} is
\begin{equation}
\begin{aligned}
\label{4.47}
\Tilde{\Phi}'+\frac{a'}{a}\Tilde\Phi= \sqrt{B^2+\frac{k^2}{3}}\; \Tilde\Phi \, 
\end{aligned}
\end{equation}
and defines the peculiar velocity.

If we neglect the contribution of the peculiar velocity into equation \rf{2.8} and consequently drop off the expression \rf{4.47}, then we get
\begin{equation}
\begin{aligned}
\label{4.48}
\frac{\Tilde{\delta A_{\mathtt{st}}}}{a^2}=
-\frac{2}{\kappa a^2}\left(k^2+\frac{a^2}{\lambda_{\mathtt{st}}^2}\right)\Tilde\Phi\equiv -\frac{2}{\kappa a^2}f_1(k) \Tilde\Phi\, ,
\end{aligned}
\end{equation}
where we took into account that the energy density fluctuations \rf{2.13} in the considered case is
\begin{equation}
\begin{aligned}
\label{4.49}
\Tilde{\delta\varepsilon}= \frac{\Tilde{\delta A_{\mathtt{st}}}}{a^2}+2\frac{\bar A_{\mathtt{st}}}{a^2}\Tilde\Phi\,  .
\end{aligned}
\end{equation}
We introduced the screening length
\begin{equation}
\begin{aligned}
\label{4.50}
\lambda_{\mathtt{st}}^2\equiv \left[\frac{\kappa \bar A_{\mathtt{st}}}{a^2}\right]^{-1}\, .
\end{aligned}
\end{equation}
It can be easily seen that
\begin{equation}
\begin{aligned}
\label{4.51}
\frac{a^2}{\lambda_{\mathtt{st}}^2}= 3B^2= \const
\end{aligned}
\end{equation}
and $\lambda_{\mathtt{s}}\equiv |u_{\mathtt{s}}|H^{-1}c =\lambda_{\mathtt{st}}$.

Obviously, Eq. \rf{4.48} is pure Helmholtz equation. However, if we preserve the velocity depended expression \rf{4.47}  in Eq.\ \rf{2.8}, then an additional $k$-depended term comes into play:
\begin{equation}
\begin{aligned}
\label{4.52} 
	\frac{\Tilde{\delta A_{\mathtt{st}}}}{a^2}
	&=-\frac{2}{\kappa a^2}\left(k^2+3B\sqrt{B^2+\frac{k^2}{3}}+\frac{a^2}{\lambda_{\mathtt{st}}^2}\right)\Tilde\Phi \\[2mm]
	&\equiv -\frac{2}{\kappa a^2}f_2(k) \Tilde\Phi\
\, ,
\end{aligned}
\end{equation}
corrupting the Helmholtz equation.

Let us consider now the model where the matter fluctuation is a localized inhomogeneity: 
\begin{equation}
\begin{aligned}
\label{4.53}
\frac{\delta A_{\mathtt{st}}}{a^2} = \frac{Mc^2 \delta(\textbf{r})}{a^3}\, 
\end{aligned}
\end{equation}

It can be easily seen from \rf{4.48} and \rf{4.52}, that, similar to the case of radiation, at short distances from the inhomogeneity (i.e., large $k$ limit), the gravitational potentials have Newtonian behavior. On the other hand, at large distances (e.g., outside the horizon $a/k\gg \lambda_{\mathtt{st}}$)
\begin{equation}
\begin{aligned}
\label{4.54}
f_2(k)\approx k^2 + 2\frac{a^2}{\lambda_{\mathtt{st}}^2}\, .
\end{aligned}
\end{equation}
Therefore, both gravitational potentials exponentially decrease at large distances. However, the consideration of peculiar velocities leads to a change in the screening length. 

The gravitational potentials in the position space are given by Eq.\ \rf{3.38} where the functions $f_i(k)$ are defined in Eqs.\ \rf{4.48} and \rf{4.52}. These integrals depend on the parameter $B$.

In Figure \rf{fig.3}, we present dimensionless potentials for a particular case $B=0.1$. The  difference between the two lines shows the effect of peculiar velocities.

We can also construct the $B$-independent dimensionless potentials as
\begin{equation}
\begin{aligned}
\label{4.55}
\phi_i(\xi)\equiv -\frac{c^2 a }{G_N M B}\Phi_i=
\frac{1}{\xi}\frac{2}{\pi}
\int_0^{\infty}dl\; \frac{l\sin(l\xi)}{\widehat{f}_i(l)}
\, , \quad i=1,2\, ,
\end{aligned}
\end{equation}
where $k/B\equiv l,\, Br\equiv \xi$ and
\begin{align}
\label{4.56}
	\widehat{f}_1(l) &\equiv l^2 +3\, ,\\
\label{4.57}
	\widehat{f}_2(l) &\equiv l^2 +3\sqrt{1+\frac{l^2}{3}}+3\, .
\end{align}
The corresponding potentials are depicted in Figure \rf{fig.4}. 

\vspace{5mm}

\begin{figure}[!ht]
	\centering
	
	\includegraphics[width=.9\linewidth]{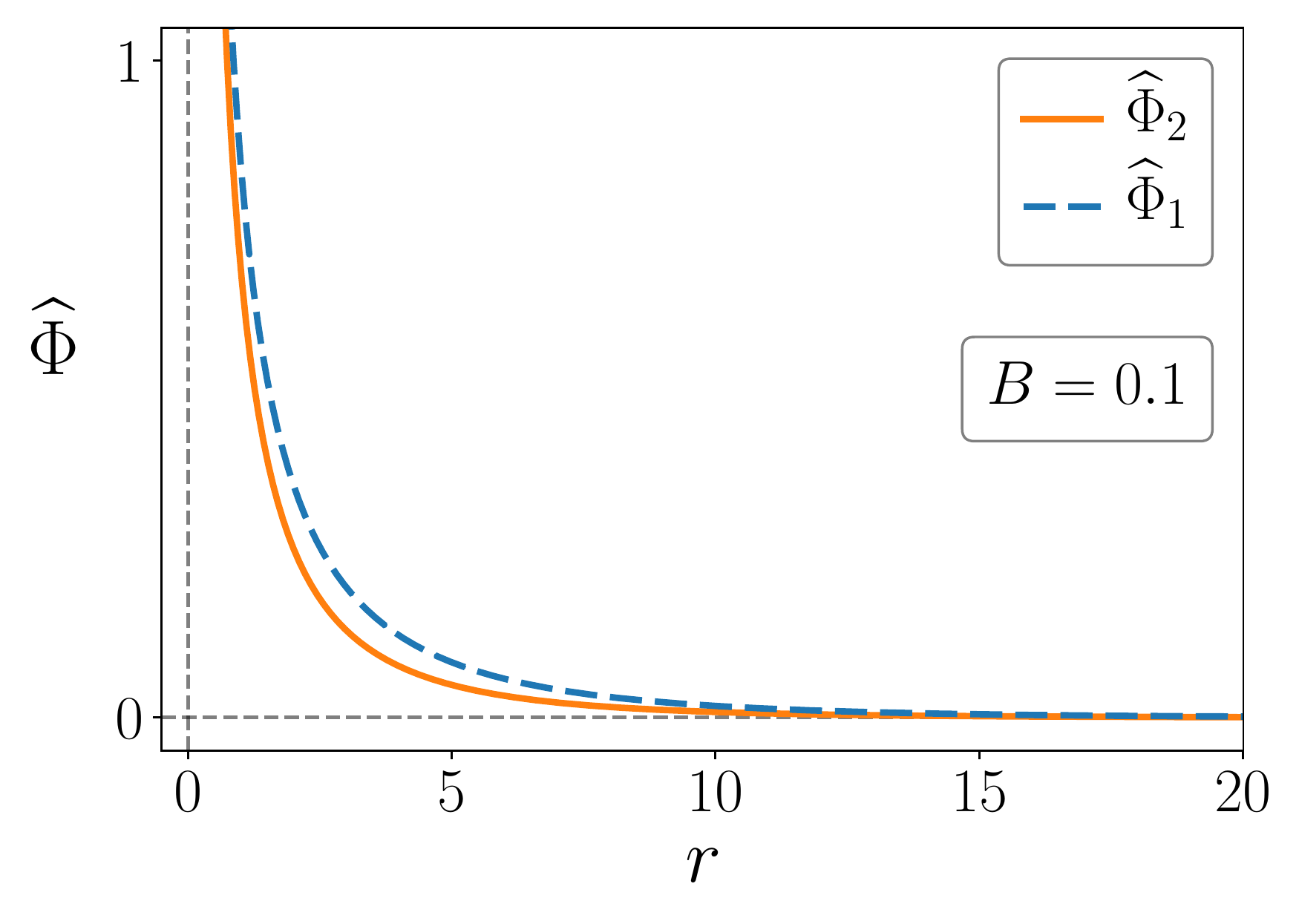}
	
	\vspace{-3mm} 
	
	\caption{Gravitational potentials in the case of the frustrated network of cosmic strings for parameter $B=0.1$. The dashed blue line corresponds to pure Yukawa potential, and the solid orange line takes into account the effect of peculiar velocity.}
	\label{fig.3}
\end{figure}


\begin{figure}[!ht]
	\centering
	
	\includegraphics[width=.9\linewidth]{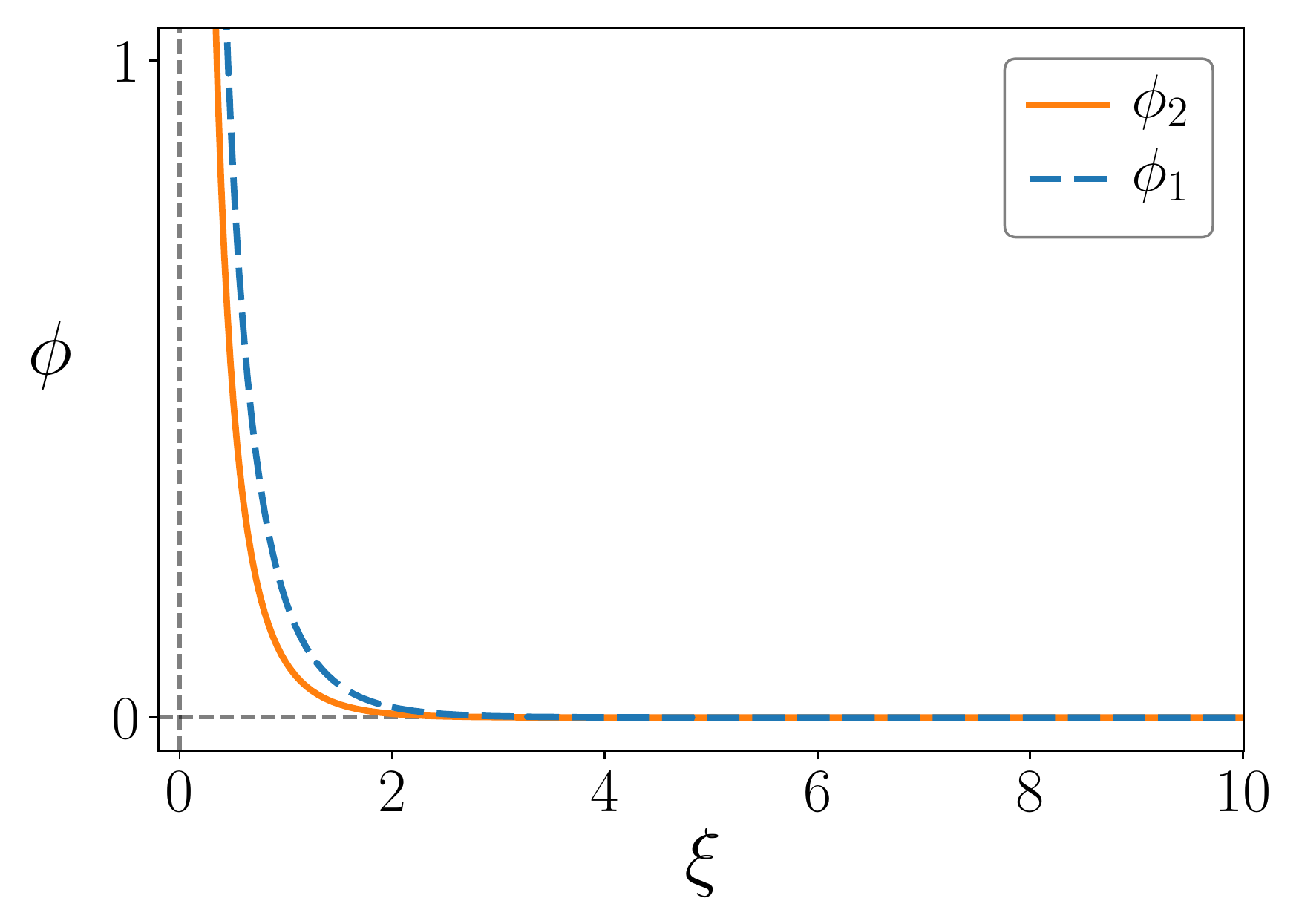} \\
	
	\vspace{-3mm}
	
	\caption{Gravitational potentials \rf{4.55} $\phi (\xi)$ where $\xi=Br$. The dashed blue and the solid orange lines have the same meaning as in Figure \rf{fig.3}.}
	\label{fig.4}
\end{figure} 

\


\section{Conclusion}

In the present paper, we have investigated the effect of the peculiar velocities on the form of the gravitational potential in cosmological models with perfect fluids. We have considered perfect fluids with the constant parameter $\omega$ of the equation of state.  Starting from an arbitrary value of $\omega$, we then concentrated on relativistic fluid with $\omega=1/3$.  Here, peculiar velocities undergo acoustic oscillations \cite{Rubakov}. In the momentum space, we have obtained the formulas for the gravitational potentials both in the presence and absence of peculiar velocities.  To get the exact form of potentials in the position space, we have assumed that the matter 
fluctuation is a localized inhomogeneity in the form of the delta function. If we neglect peculiar velocities, then the gravitational potential has the form of the Yukawa potential. 
Since the Fourier integral for the velocity-dependent potential can be calculated only numerically, we have depicted the results graphically in Figures \rf{fig.1} and \rf{fig.2}. These figures clearly demonstrate  the modulation of the gravitational potential by acoustic oscillations due to the presence of peculiar velocities. 

To illustrate the effect of the peculiar velocities on the gravitational potential, we also considered the case of the frustrated network of cosmic strings with $\omega =-1/3$. The result is depicted in Figures \rf{fig.3} and \rf{fig.4}. In this exceptional case, acoustic oscillations are absent. Nevertheless, the difference between the figures demonstrates the effect of peculiar velocity.


\section*{Acknowledgements}

We thank Maxim Eingorn for helpful discussions.


\end{document}